\documentclass[12pt]{article}
\usepackage{graphicx}
\usepackage{amssymb}
\usepackage{amsmath}
\usepackage{amsthm}
\usepackage{epsf}
\usepackage[parfill]{parskip}
\usepackage[matrix,arrow]{xy}
\input{epsf}

\newtheorem*{theorem}{Theorem}

\newcommand{\ba}{\begin{eqnarray*}}
\newcommand{\ea}{\end{eqnarray*}}
\newcommand{\ban}{\begin{eqnarray}}
\newcommand{\ean}{\end{eqnarray}}

\newcommand{\IZ}{\mathbb{Z}}
\newcommand{\IC}{\mathbb{C}}

\newcommand{\IR}{\mathbb{R}}

\newcommand{\cM}{{\cal M}}

\newcommand{\cN}{{\cal N}}
\newcommand{\cV}{{\cal V}}

\newcommand{\tp}{\tilde{p}}
\newcommand{\tg}{\tilde{g}}
\newcommand{\tomega}{\tilde{\omega}}

\newcommand{\re}{{\rm Re \,}}
\newcommand{\im}{{\rm Im \,}}

\def \nn{\nonumber}

\newcounter{n}

\makeatletter
\@addtoreset{equation}{section}
\makeatother

\begin{document}
\noindent
\begin{titlepage}

\begin{center}
\hfill ITFA-2006-42\\
\hfill SPHT-T06/147\\
\hfill hep-th/0611106\\ 
\vskip 2cm {\Huge
Towards reduction of\\ \vskip 0.1cm
type II theories \\ \vskip 0.3cm
on $SU(3)$ structure manifolds}
\vskip 1cm {Amir-Kian Kashani-Poor$^1$ and Ruben Minasian$^2$}\\ \vskip 0.5cm
{\it $^1$Institute for Theoretical Physics, University of Amsterdam\\
1018 XE Amsterdam, The Netherlands\\} \vskip0.2cm
{\it $^2$Service de Physique Th\'eorique, CEA/Saclay\\
91191 Gif-sur-Yvette Cedex, France}
\end{center} 
\vskip 1.5cm

\begin{abstract}
We revisit the reduction of type II supergravity on $SU(3)$ structure manifolds,
conjectured to lead to gauged $\cN=2$ supergravity in 4 dimensions. The reduction proceeds by expanding the invariant 2- and 3-forms of the $SU(3)$ structure as well as the gauge potentials of the type II theory in the same set of forms, the analogues of harmonic forms in the case of Calabi-Yau reductions. By focussing on the metric sector, we arrive at a list of constraints these expansion forms should satisfy to yield a base point independent reduction. Identifying these constraints is a first step towards a first-principles reduction of type II on $SU(3)$ structure manifolds.
\end{abstract}
\end{titlepage}
\newpage

\section{Introduction}
Taking a higher dimensional theory as a starting point, more than one path can lead to a lower dimensional theory. The conventional and most physical connection is obtained by Kaluza-Klein reduction: the starting point is chosen from a specific class of solutions (vacua) of the higher dimensional theory (essentially solutions that factorize between the dimensions one wants to keep and those one would like to discard). The lower dimensional theory is obtained by expansion around one such solution and describes light fluctuations around it. A different requirement one can impose on the lower dimensional theory, which sometimes goes under the name of non-linear reduction, is that its solutions lift to solutions of the higher dimensional theory. The reduction of 11d SUGRA on topological $S^7$ to $\cN=8$ SUGRA in 4d \cite{dWN} is a prominent example of such a relation between a higher and lower dimensional theory. Note that it is not guaranteed that such an ansatz captures all light degrees of freedom around each of the incorporated higher dimensional solutions.

Where is the familiar Calabi-Yau reduction of type II theories \cite{CHSW,BCF} situated with regard to these two possibilities? The reduction can be performed by choosing a Ricci flat metric ${\bf g_0}$ on the Calabi-Yau $X$, and expanding the fields in terms of ${\bf g_0}$-harmonic forms $\omega_i$. We will refer to this in the following as a base point dependent reduction, since we are expanding around a solution ${\bf g_0}$, the hallmark of a Kaluza-Klein reduction. However, we generically have a continuous family of solutions ${\bf g}(t)$, and we can free our ansatz from the base point dependence on ${\bf g_0}$ by expanding in ${\mathbf g}(t)$-harmonics $\omega_i(t)$ instead. The $t$ are metric moduli, and one might hence expect the reduction of the metric sector of the theory to be significantly modified by this step $\omega_i \rightarrow \omega_i(t)$. This does not happen, as we will review below, as the 4d theory ends up depending only on the cohomology classes of the forms $\omega_i(t)$ \cite{Strominger, CdlO, s2}, which of course do not vary with $t$.

Mainly due to this latter fact, the reduction can be performed without having an explicit expression for the expansion forms (a lucky circumstance, since no Ricci flat metrics on compact Calabi-Yau manifolds are explicitly known, let alone explicit expressions for harmonic forms). The 4d theory is expressed in terms of some topological and holomorphic data of the Calabi-Yau (the triple intersection number of the 2nd cohomology and the period matrices of the complex structure). This data is precisely what is needed to specify an (ungauged) $\cN=2$ supergravity action in 4d, and organizes itself appropriately upon performing the reduction.

Above, the requirement we imposed on a non-linear reduction was that the solutions of the lower dimensional theory lift to solutions of the higher dimensional theory. How does the Calabi-Yau reduction fare on this account? Since the ungauged 4 dimensional $\cN=2$ action does not exhibit a potential term, all constant values for the scalar fields are a solution to the 4d equation of motion, and by construction lift to solutions of the higher dimensional theory. While no proof of this lifting property for arbitrary solutions exists to our knowledge, it does hold for certain other prominent solutions such as $\cN=2$ black holes \cite{HMT}.

Flux compactifications establish a connection between string theory and {\it gauged}  $\cN=2$ supergravity. Indeed, as first shown in \cite{PS}, nonvanishing expectation
values of the internal fluxes are described in the 4d effective theory by the scalars in the hypermultiplets picking up charges under the gauge fields in the vector multiplets. The fluxes contribute to the potential of the 10d theory, and this energy is reproduced correctly by the potential term in gauged supergravity. The reduction in the presence of fluxes
is still performed on a Calabi-Yau manifold \cite{M,CKLT,DA,LoMi,KK}, and the resulting theory has the same spectrum as its flux-less relative. In particular, it is based on expanding fields in the harmonic forms on the internal Calabi-Yau. The justification for this procedure is still not established (but see \cite{vafa, LMc}). Note that a Kaluza-Klein reduction would take the backreacted geometry as a starting point and would yield a 4d effective theory, generically non-supersymmetric, valid around a given VEV of the 4d scalar fields. The hope is that the procedure described above yields an effective theory encompassing multiple solutions of the 10d theory at different minima of its potential.\footnote{In fact, merely turning on fluxes can never result in a potential with minima at finite radius, as the contribution of fluxes to the potential energy is minimized when the fluxes are `diluted' in the decompactification limit. See \cite{KK} for one possibility to avoid this runaway behavior in the effective $\cN=2$ context.}

Ignoring for the moment the various conceptual challenges posed by effective $\cN=2$ descriptions of flux compactifications, one can consider gauged $\cN=2$ theories in light of the swampland program \cite{VafaSL}: having obtained an $\cN=2$ theory from compactification, can all of its possible gaugings be realized within string theory? Flux compactifications do not exhaust all possible gauging. Recently, various authors \cite{GLMW, AFTV, GLW, HP,LM} have suggested that gauged $\cN=2$ supergravity can also be obtained by compactifying on $SU(3)$ structure manifolds. These manifolds admit almost complex and symplectic structures and hence possess invariant forms $J$ and $\Omega$ (and nowhere vanishing spinors) just as Calabi-Yau manifolds do, but these structures are no longer required to be integrable. When considered as deformations of Calabi-Yau geometries (see e.g. \cite{T}), these ans\"atze supply the missing gaugings \cite{AFTV}. Once these reductions are better understood, however, they should be able to stand on their own feet (the swampland question then of course would arise in the opposite direction, possibly indicating that these manifolds can {\it always} be understood as deformations of Calabi-Yau manifolds).

The reduction proceeds by mimicking the ansatz for Calabi-Yau reductions. In the latter, the expansion forms $\omega_i(t)$ are specified geometrically (as harmonic forms) and their relation to the moduli space of Calabi-Yau metrics is known. That these forms satisfy all the properties needed for the reduction to go through and the 4d action to assemble itself into $\cN=2$ supergravity hence is required by consistency. By contrast, the space of metrics that should be considered in the more general $SU(3)$ structure case is not well understood. The procedure in the literature, which we shall follow and review in much greater detail below, has therefore been the following: to allow for manifolds with merely $SU(3)$ structure rather than $SU(3)$ holonomy, we must allow for some expansion forms to be non-closed. We then attempt to impose the minimal number of requirements on such a system of forms for the resulting four-dimensional theory to have the structure required by $\cN=2$ supersymmetry. 

The starting point for the analysis in this note is the above observation that in the case of CY reductions, the step from base dependent to base independent reduction, $\omega_i \rightarrow \omega_i(t)$, is unproblematic due to the reduction depending only on the cohomology classes of the forms $\omega_i(t)$. In the modified setup, such considerations do not apply (the expansion forms are not closed). As explained above, problems are expected to arise in the metric sector, and we hence expose the reduction of this sector to more scrutiny than has been hitherto done. We have essentially two results to report: for the base point dependent reduction to go through, certain differential conditions must be satisfied by the forms, but we demonstrate that these are equivalent to conditions that have been assumed to hold already. We find one additional constraint which is new and must be imposed. The step to a base point independent reduction requires imposing additional constraints, which we discuss. The constraints appear very restrictive.

Throughout this paper, we present our results in the framework of type IIA.

We begin in section \ref{s:theforms} by reviewing and completing the conditions that have appeared in the literature on the system of forms the reduction is to be based on, and listing the additional conditions needed for a base point independent reduction. In section \ref{s:rms}, we analyse the reduction of the metric sector of the theory. We derive the conditions for the base point dependent reduction to go through and see that these follow from the conditions imposed in section \ref{s:theforms}. We also demonstrate how the conditions for the base point independent reduction arise. In section \ref{s:eeL}, we clarify the relation of our ansatz to one based on expanding in eigenforms of the Laplacian. We construct a system of forms satisfying the na\"{\i}ve conditions required for the reduction to go through, and discuss its shortcomings.

\section{Conditions on the expansion forms}  \label{s:theforms}
The starting point of the analysis is a reduction manifold $X$ which has $SU(3)$ structure, but is not necessarily Calabi-Yau. Such manifolds exhibit a set of $SU(3)$ invariant forms, a 2-form $J$ and a 3-form $\Omega$. As the nomenclature indicates, these will play a similar role in the reduction as the K\"ahler form and the holomorphic 3-form do in Calabi-Yau reductions. In particular, $J$ determines an $Sp(6,\IR)$ structure, and $\Omega$ an $SL(3,\IC)$ structure. As in Calabi-Yau reductions, $J$ and $\Omega$ are to be expanded in the same set of forms as the RR gauge potentials and the B-field,
\ban
J = v^i \omega_i \, , \qquad \Omega= X^A \alpha_A - G_A \beta^A \,. \label{expinvf}
\ean

Let us recall that $J$ and $\Omega$ are no longer closed, and their failure to be such (i.e. the failure of the structure group to be the holonomy group) is characterized by components of the intrinsic torsion, which fit into SU(3) representations,
\ban
dJ &=& -\frac{3}{2}\, {\rm Im}(W_1 \bar{\Omega}) + W_4 \wedge J + W_3 \, ,  \nn \\
d\Omega &=& W_1 J^2 + W_2 \wedge J + \bar{W_5} \wedge \Omega\ \, .   
\ean
It follows that the expansion forms cannot all be closed, and we must choose what conditions to impose on their differentials. The smallest deviation from the Calabi-Yau reduction, while allowing for non-closed $J$ and $\Omega$, is given by the following ansatz.
\begin{enumerate}
\item{We start with a set of 2-forms $\omega_i$.} \label{e2f}
\item{We need a set of dual 4-forms $\tomega^i$ such that 
\ban
\int \omega_i \wedge \tomega^j &=& \delta_i{}^j \,.  \label{24dual}
\ean
For a Calabi-Yau, these exist by Poincar\'e duality. Here, we construct them by requiring the matrix
\ba
g_{ij} = \int \omega_i \wedge *\omega_j  \,,
\ea
to be invertible with inverse $g^{ij}$, and defining
\ban
\tilde{\omega}^i = g^{ij} * \omega_j \,. \label{defining4}
\ean} \label{d4f}
\item{The 3-forms are to come in pairs $\alpha_A, \beta^A$ and should satisfy
\ban
\int \alpha_A \wedge \beta^B &=& \delta_{A}{}^{B} \,,\nn\\
\int \alpha_A \wedge \alpha_B &=& \int \beta^A \wedge \beta^B =0 \,. \label{symplbasis}
\ean
In addition, the Hodge duals of this set of 3-forms should be expressible as linear combinations within the same set,
\ban
* \alpha_A &=& A_A^B \alpha_B + B_{AB} \beta^B \,,  \nn \\
*\beta^A &=& C^{AB} \alpha_B - A^A_B \beta^B \,,  \label{hodge}
\ean
with constant (i.e. coordinate independent) coefficient matrices $A, B, C$.}  \label{p3f}
\item{For the variation of the coefficients of the $\alpha_A$ in the expansion
\ba
\Omega= X^A \alpha_A - G_A \beta^A
\ea
to correspond to variations of the $SL(3,\IC)$ structure, we must require that the forms
\ban
 \alpha_A - \partial_A G_B \beta^B -\kappa_A \Omega \label{bidegreeconstr}
\ean
be of type (2,1) away from the $X^A=0$ locus. The objects that enter in the definition of these forms are introduced in section \ref{fcs}.}  \label{vcs}
\item{The most obvious differential constraints to impose are that the set of 2-, 3-, and 4-forms we expand in are closed under the action of $d$ and $d^\dagger$. This yields \cite{GLMW, AFTV, GLW}
\ban
d^\dagger \omega_i &=& 0 \label{ddagom} \\
d\omega_i &=& m_i{}^A \alpha_A + e_{i A} \beta^A \label{exp2forms} \\
d \alpha_A = e_{iA} \tomega^i &;& d\beta^A = -m_i{}^A \tomega^i \label{exp3forms} \\
d\tomega^i &=& 0  \,.  \label{dc4f}
\ean 
Note that under the assumption of closure under the action of $d$, $d^\dagger$, this is the most general set of conditions we can impose (the coefficients in (\ref{exp3forms}) follow from (\ref{24dual}) and (\ref{symplbasis})). For consistency ($d^2 =0$), the coefficient matrices must satisfy the following set of constraints
\ban
m_i{}^A e_{jA} - e_{iA} m_j{}^A =0  \,.   \label{nilpcond}
\ean
Upon performing the reduction with such an ansatz, the matrices $m_i{}^A$ and $e_{iA}$ descend to charge matrices for the hypermultiplets under the vectors. We hence require that they have integer entries.} \label{d23f}
\item{We next need to impose conditions on the forms
\ba
A_{iA} = \omega_i \wedge \alpha_A &;& B_i{}^A= \omega_i \wedge \beta^A \,.
\ea
The need for constraints on these forms is apparent at many points in the reduction. The strongest constraints, from which all others follow, arise from our analysis in section \ref{sec:sg}, and are given by
\ban
X^A A_{iA} - G_A B_i{}^A &=& 0 \,, \label{omega11}\\
v^i (mA +eB)_{(ij)} &=& 0 \,.  \label{con2AB}
\ean
Note in particular that (\ref{omega11}) is just the condition $\omega_i \wedge \Omega =0$, hence implies that the 2-forms $\omega_i$ are of type (1,1), and the 4-forms $\tomega^i$ consequently of type (2,2). This condition also implies compatibility of $J$ and $\Omega$,
$J \wedge \Omega = 0$.
}  \label{comp}
\setcounter{n}{\theenumi}
\end{enumerate}
By imposing the conditions \ref{e2f} through \ref{comp} (excluding \ref{vcs}, the need for which will become apparent in the following section), it has been shown \cite{GLMW, AFTV, LM} that the reduction of the terms in the 10d action involving the RR and NSNS gauge potentials yield the expressions familiar from Calabi-Yau reductions, but with the derivatives acting upon the hyperscalars elevated to gauge covariant derivatives, with the charges of the scalars being dictated by the integer entries of the coefficient matrices $e_{iA}$ and $m_i{}^A$. Furthermore, additional terms from these sectors not present in conventional Calabi-Yau reductions assemble themselves, together with the terms stemming from the reduction of $R_6$, to the potential of $\cN=2$ gauged supergravity dictated by the charges of the hyperscalars. That the reduction of $R_6$ yields the correct terms has been shown \cite{GLMW, LM} under the assumptions that the components of the intrinsic torsion in the
representations $\bf 3$ and $\bf {\bar 3}$ vanish, i.e. $J\wedge dJ =0$ and $d\Omega^{(3,1)}=0$, hence $W_4=W_5=0$.\footnote{When $W_4=W_5=0$, the
internal Ricci scalar can be written as \cite{BV}
\ba
R_6 = \frac{1}{2} (15 W_1 {\bar W_1} - {W_2} \llcorner {\bar W_2} - W_3
\llcorner W_3)\,,
\ea
where on forms of any degree, $W \wedge * W = (W \llcorner W) {\rm
Vol}_6$. Introducing pure spinors $\Phi_+ = \exp(-iJ)$ and $\Phi_- =
\Omega$, we observe that the structure of $R_6$ is matched by
\ba
\frac{1}{2} \left(\langle d \Phi_+ , * d {\bar \Phi}_+ \rangle + \langle d
\Phi_- , * d {\bar \Phi}_- \rangle \right) \,,
\ea
where we have used the standard definition of the Mukai pairing $ \langle
\cdot, \cdot \rangle$,
see e.g. \cite{GLW}. This contribution to the
potential would nicely combine with that of the NS flux into
\ba V_{\rm NS}
= \frac{1}{2}\left( ( dJ+iH) \wedge * (dJ-iH) + d \Omega \wedge *d {\bar
\Omega}\right) \,, 
\ea
which has the mirror-symmetric structure advocated in \cite{GLMW, FMT, GMPT1}.}
 These conditions follow from (\ref{con2AB}) and (\ref{omega11}), respectively. Condition \ref{vcs} has not been discussed in the literature previously. 

In the following section, we perform the reduction of the metric sector. We will see that the conditions listed above are sufficient for the reduction to work if we assume that the expansion 2- and 3-forms do not vary with the metric moduli (by definition (\ref{defining4}), the 4-forms $\{\tilde{\omega}^i\}$ are moduli dependent even for a fixed choice of 2-forms $\{\omega_i\}$). If we instead allow such a variation (recall that in the Calabi-Yau case, we expand in harmonic forms that hence are moduli dependent), we need to impose further conditions on these variations. 

To retain the form of the prepotential in the vector multiplet sector, as expressed in terms of the forms $\{\omega_i \}$, upon allowing these forms to depend on the moduli, and likewise to retain the form of the special geometry part of the quaternionic metric, the following three conditions arise.

\renewcommand{\labelenumi}{$*$\arabic{enumi}.}

\begin{enumerate}
\setcounter{enumi}{\then}
\item{The 2-forms should satisfy the constraint
\ba
v^i \frac{\partial}{\partial v^j} \omega_i &=& 0  \,,
\ea
with the $v^i$ metric moduli as defined in (\ref{expinvf}). We will review why this holds in the Calabi-Yau case in the next section.} \label{headache}
\item{The integral
\ban
d_{ijk} &=& \int_X \omega_i \wedge \omega_j \wedge \omega_j \label{tripleintersection}
\ean
should be moduli independent. In the Calabi-Yau case, this is guaranteed because the derivative of the harmonic form $\omega_i$ with regard to a metric modulus is exact.\footnote{Note that we are not requiring $d_{ijk}$ to be a topological invariant. E.g., it can depend on geometric data specifying the subset of $SU(3)$ structures encompassed by our parametrization.}} \label{triple}
\item{Analogously, we demand the vanishing of the following integrals,
\ban 
\int \alpha_A \wedge \partial_C \alpha_B =\int \alpha_A \wedge \partial_C \beta^B= \int \beta^A \wedge \partial_C \beta^B=0 \,, \label{inlieuexact}
\ean 
where the derivatives are taken with regard to metric moduli that will be introduced in subsection \ref{fms}. Again, the vanishing of these integrals is guaranteed in the Calabi-Yau case by the exactness of the derivatives.} \label{h3f}
\end{enumerate}
We have labelled these final three conditions with a $*$, as they are derived under the assumption that we retain the form of the prepotentials after allowing moduli dependence of the expansion forms. Can this assumption be weakened? It is possible that the correct reduction requires adding contributions to the prepotentials which depend on derivatives of the expansion forms and hence vanish in the case that these are constant. Though we have not been able to come up with such an ansatz, we are not claiming a no-go theorem in this direction. 

Basing the reduction on constant expansion forms is the analogue of picking a base point in moduli space in the case of Calabi-Yau reductions, and expanding in forms harmonic with regard to the metric at this point. This vantage point makes do with the requirements \ref{e2f} to \ref{comp}. Such an ansatz however does not seem in keeping with the underlying philosophy of the reduction, that it be valid over all of moduli space. Removing the base point dependence necessitates imposing additional conditions on the forms. The most natural choice appears to be conditions $*$\ref{headache} to $*$\ref{h3f}.

\section{Reduction of the metric sector} \label{s:rms}
\subsection{Special geometry} \label{sec:sg}
Vector fields arise from the expansion of the RR 3-form field $C_3$ in the set of 2-forms $\{\omega_i \}$. By $\cN=2$ supersymmetry, these vectors should be accompanied by complex scalars, parametrizing a scalar manifold with a special K\"ahler metric. In analogy to the Calabi-Yau case, these scalars should arise in our compactification scheme from the variation of the $Sp(6,\IR)$ structure. 

Let us briefly review the Calabi-Yau case. We start by specifying a basis $\{\Gamma^i\}$ of $H_2(X,\IZ)$. Coordinates $v^i$ on the space of K\"ahler classes are then introduced via
\ba
v^i &=& \int_{\Gamma^i} J  \,,
\ea
for $J$ an arbitrary representative of the K\"ahler class $[J]$. By Yau's theorem, given a complex structure on $X$ and the K\"ahler class specified by the $v^i$, we can find a Ricci flat metric with associated K\"ahler form $J$ within this K\"ahler class. Hence, $v$ not only specifies a K\"ahler class but also a K\"ahler form, which we will denote by $J(v)$.
A K\"ahler form together with a complex structure on $X$ uniquely determine a metric via
\ba
ig_{a \bar{b}} = J_{a\bar{b}} \,.
\ea
To consider variations of this metric with regard to the coordinates $v^i$, we introduce a basis $\{[\omega_i]\}$ of integral cohomology $H^2(X,\IZ)$ dual to the basis $\{ \Gamma^i \}$ introduced above, with the $\omega_i(v)$ representatives that are harmonic with regard to the metric determined by $J(v)$. Then,
\ban
i \frac{\partial g_{a \bar{b}}}{\partial v^i} = \omega_{i\;a\bar{b}} +  v^j \frac{\partial}{\partial v^i}\omega_{j\;a\bar{b}}  \,. \label{metricvariation}
\ean
By the Lichnerowicz equation, we know that variations of a Ricci flat metric preserve Ricci flatness if and only if the associated 2-form
\ba
\frac{\partial g_{a \bar{b}}}{\partial v^i} dz^a \wedge d\bar{z}^{\bar{b}}
\ea
is harmonic. Of the forms appearing on the RHS of (\ref{metricvariation}), $\omega_i$ is harmonic by definition. $\partial_i \omega_j$ is exact, as $[\omega_i(v)]$ is constant, hence we can conclude
\ba
v^j \frac{\partial}{\partial v^i}\omega_{j\;a\bar{b}}  = 0\,.
\ea
At our current understanding of the $SU(3)$ structure case, we must skip several of the steps above, and take as our starting point an $SU(3)$ invariant 2-form $J$ together with ad hoc coordinates $v^i$ on the correct subspace of $Sp(6,\IR)$ structures such that $J= v^i \omega_i(v)$.
 
Using the $SU(3)$ invariant form $\Omega$ to introduce, patchwise, a basis of $T^*X$ of definite type, we can then define a hermitian metric on $X$ in terms of $J$ as
\ba
ig_{a \bar{b}} = J_{a\bar{b}} 
\ea
and consider its variation with regard to $v^i$,
\ba
i \frac{\partial g_{a \bar{b}}}{\partial v^i} = \omega_{i\;a\bar{b}} +  v^j \frac{\partial}{\partial v^i}\omega_{j\;a\bar{b}}  \,.
\ea
With this relation, KK reduction of the Ricci scalar $R$ yields the following metric for the $\sigma$-model describing the almost symplectic sector,
\ban
\cV G_{i j} (v)&\sim& (\delta_i{}^k + v^k \frac{\partial}{\partial \tilde{v}^i} ) |_{\tilde{v} = v}\, (\delta_j{}^l + v^l \frac{\partial}{\partial v'^j} ) |_{v' = v}  \int_X \omega_k(\tilde{v}) \wedge * \omega_l(v') \,,  \label{symplmetric}
\ean
where we have introduced $\cV = \int J \wedge J \wedge J$. The Hodge star is taken with regard to the metric $g_{a \bar{b}}(v)$. It can be traced back to the contractions required to obtain the Ricci scalar from the Riemann tensor, as in the Calabi-Yau case \cite{BCF}. 

For Calabi-Yau reductions, a crucial ingredient in obtaining special geometry from the reduction of the symplectic sector is the complexification of the $v^i$ by the scalars $b^i$ descending from the expansion of the NSNS $B$-field, $B= b^i \omega_i + \ldots$, to $t^i = b^i + i v^i$. The kinetic term for these scalars arises from the reduction of $\int_X H \wedge *H$, hence has $\sigma$-model metric
\ban
G^B_{i j} (v) &\sim& \frac{1}{\cV} \int_X \omega_i(v) \wedge * \omega_j(v) \,. \label{gb}
\ean
Clearly, $G^B$ must coincide with the metric in (\ref{symplmetric}) for this complexification to take place.\footnote{Under our general assumption that the functional form of the prepotential is not modified upon admitting moduli dependence of the expansion forms, we can argue that the complexification $t^i = b^i + i v^i$ must take place in precisely this form by considering the gauge sector.} The derivative terms in (\ref{symplmetric}) must hence vanish. By considering the diagonal contribution
\ba
v^k \frac{\partial}{\partial \tilde{v}^i}  |_{\tilde{v} = v}\,  v^l \frac{\partial}{\partial v'^i}  |_{v' = v}  \int_X \omega_k(\tilde{v}) \wedge * \omega_l(v') &=& || v^k \frac{\partial}{\partial v^i} \omega_k(v) ||^2 \,,
\ea
we recognize that short of miraculous cancellations between various integrals, we must require $v^k \frac{\partial}{\partial v^i} \omega_k(v) = 0$. This is our condition $*$\ref{headache}, and with it, (\ref{symplmetric}) reduces to (\ref{gb}), and we can henceforth drop the ${}^B$ in referring to this metric. 

The expression for $G$ can be considerably simplified to reveal the special geometry underlying it, {\it provided we assume the expansion forms $\omega_i$ are of type (1,1)}. This is where the need for condition (\ref{omega11}) arises. Let us begin by reexpressing $*\omega_i$. Given an almost complex structure on $X$ with regard to which $\omega_i$ is of type $(1,1)$, we consider a patch and introduce local complex coordinates $z^{\alpha}$, inducing a basis of definite type for the cotangent space. Furthermore, we can choose this basis so that {\it at a point $P_0$}, the $SU(3)$ invariant 2-form $J = \frac{i}{2}\sum dz^{\alpha} \wedge d\bar{z}^{\bar{\alpha}}$. A purely algebraic calculation now yields \cite{Strominger}, at $P_0$,
\ban
* \omega_i &=& \frac{1}{2} ( \omega_i \llcorner J) J \wedge J -  \omega_i \wedge J  \,.  \label{staromega}
\ean
This equality extends to the whole patch, as it is formulated intrinsically (without reference to the point $P_0$). To extend it over all of $X$, we need $J$ to be a globally defined nowhere vanishing $(1,1)$ form which at a given point can be put in the diagonalized form. $J$ of course enjoys these properties courtesy of the $SU(3)$ structure we take as our starting point.

Next, we want to reexpress the contraction $\omega_i \llcorner J$. Consider
\ba
\frac{1}{2} \int_X ( \omega_i \llcorner J) J \wedge J \wedge J &=& \int_X * \omega_i \wedge J + \int_X \omega_i \wedge J \wedge J \\
&=&  \int_X  \omega_i \wedge *J +  \int_X \omega_i \wedge J \wedge J \\
&=& \frac{3}{2}\int_X \omega_i \wedge J \wedge J \,.
\ea
To pull $\omega_i \llcorner J$ out from underneath the integral, we need $d(\omega \llcorner J)=0$. But this is a consequence of (\ref{dc4f}), {\it assuming that $d(\omega_i \wedge J)=0$},
\ba
0 &=& d \tomega^i  \\
&=& g^{ij} d * \omega_j \\
&=& g^{ij} d (\frac{1}{2}( \omega_j \llcorner J) J \wedge J -  \omega_j \wedge J) \\
&=& \frac{1}{2}g^{ij}  d( \omega_j \llcorner J) J \wedge J \,.
\ea
To ensure this relation, we have imposed condition (\ref{con2AB}). With this, we obtain the same expression for the contraction as in the Calabi-Yau case \cite{Strominger}, 
\ban
\omega_i \llcorner J&=& 3 \frac{\int_X \omega_i \wedge J \wedge J}{\int_X J \wedge J \wedge J} \,.   \label{contraction}
\ean
By plugging all this back into the expression (\ref{symplmetric}) for $G$, we see that the dependence on $\omega_i(v)$ arises in the form 
\ba
d_{ijk}(v) &=& \int_X \omega_i(v) \wedge \omega_j(v) \wedge \omega_k (v) \,.
\ea
To relate the metric $G$ to the K\"ahler form $\log K \sim \log \int J \wedge J \wedge J$, we must require that $d_{ijk}$ is independent of $v$. This is condition $*$\ref{triple}. Reexpressing the $v^i$ in terms of the complex coordinates $t^i$, we then obtain $G$ as
\ba
G_{i \bar{\jmath}} &\sim& \partial_i \partial_{\bar{\jmath}} K  \,.
\ea
Special geometry now follows exactly as in the Calabi-Yau case.

\subsection{Quaternionic geometry}  \label{fcs}
A set of 4d scalars arises when expanding the RR 3-form $C_3$ in the set of 3-forms $\{ \alpha_A, \beta^A \}$. In analogy with the Calabi-Yau case, these are to be augmented by scalars stemming from the variation of the $SL(3,\IC)$ structure. Together, these scalars are to parametrize a quaternionic manifold. We consider the metric and the RR scalars in turn.

\subsubsection{The metric scalars} \label{fms}
Let us first determine the relation between the variation of the $SU(3)$ invariant form $\Omega$ and the metric.
To this end, let $p$ be an element of the reduced $SU(3)$ frame bundle, and $\{e_a\}$ the standard holomorphic basis of $\IC^3$. Then
\ba
\Omega(p(e_a), p(e_b), p(e_c)) = \Omega_{abc}
\ea
is the invariant tensor. Now consider the infinitesimal deformation $\tilde{\Omega} = \Omega + \delta \Omega$, and let $\tilde{p}$ denote an element of the frame bundle defined by $\tilde{\Omega}$, with $\tp(e_a) = p(e_a) + \delta p_a^{\,\,b} p(e_b)$. Then
\ba
0 &=& (\Omega + \delta \Omega)(\tp(e_a), \tp(e_b), \tp(e_{\bar{c}})) \\
&=& \Omega_{abd} \delta p_{\bar{c}}^{\,\,d} + \delta \Omega_{ab\bar{c}}  \,.
\ea
Hence,
\ban
\delta p_{\bar{c}}^{\,\,d} = - \frac{1}{2||\Omega||^2} \bar{\Omega}^{abd}(\delta \Omega)_{ab \bar{c}} \,,   \label{deltap}
\ean
with $||\Omega||^2 := \frac{1}{3!} \bar{\Omega}^{abc} \Omega_{abc}$ and where we have used $\bar{\Omega}^{abc} \Omega_{abd} = \frac{1}{3} \delta^{c}{}_d \bar{\Omega}^{abe} \Omega_{abe}$.
The metric $\tilde{g}$ defined by the new structure satisfies
\ba
0 &=& \tg (\tp_{\bar{a}},\tp_{\bar{b}}) \\
&=& \delta g(p_{\bar{a}},p_{\bar{b}}) + g(p_{\bar{a}},p_c) \delta p_{\bar{b}}^{\,\,c} + g(p_c,p_{\bar{b}}) \delta p_{\bar{a}}^{\,\,c}  \,.
\ea
We thus arrive at
\ban
\delta g_{\bar{a} \bar{b}} &=& - g_{\bar{a} c} \delta p_{\bar{b}}^{\,\,c} - g_{c \bar{b}} \delta p_{\bar{a}}^{\,\,c} \nn\\
&=&  \frac{1}{2||\Omega||^2} (\bar{\Omega}^{cd}{}_{\bar{a}}(\delta \Omega)_{cd \bar{b}}+\bar{\Omega}^{cd}{}_{\bar{b}}(\delta \Omega)_{cd \bar{a}}) \nn \\
&=&  \frac{1}{||\Omega||^2} \bar{\Omega}^{cd}{}_{\bar{a}}(\delta \Omega)_{cd \bar{b}} \hspace{1cm}\mbox{for $\delta \Omega$ primitive} \,. \label{cvm}
\ean

Now assume that we have parametrized the variation of the $SL(3,\IC)$ structure in terms of parameters $z^\alpha$. Below, we will use the expansion forms $\alpha_A, \beta^A$ to define such a parametrization. Given such $z^\alpha$, we introduce 3-forms $\chi_{\alpha}$ of type (2,1) as the (2,1) part of the following derivatives,
\ban
\chi_\alpha :=\left[\frac{\partial}{\partial z^\alpha} \Omega \right]_{(2,1)}  \,.  \label{chi}
\ean
Note that $\chi_\alpha \neq 0$ by assumption of $z_\alpha$ being a parametrization of $SL(3,\IC)$ structure: two complex 3-forms that are each of type (3,0) with regard to the $SL(3,\IC)$ structure defined by the respective other 3-form define the same $SL(3,\IC)$ structure. By the compatibility condition $J \wedge \Omega =0$, the $\chi_\alpha$ are primitive.
In terms of these definitions, (\ref{cvm}) becomes
\ba
\frac{\partial}{\partial z^\alpha} g_{\bar{a} \bar{b}} &=&  \frac{1}{||\Omega||^2} \bar{\Omega}^{cd}{}_{\bar{a}}(\chi_\alpha)_{cd \bar{b}} \,. 
\ea

Reduction of the Einstein term with this ansatz yields \cite{BCF}
\ban
G_{\alpha \bar{\beta}} \sim \frac{1}{||\Omega||^2} \int_X \chi_\alpha \wedge \bar{\chi}_{\bar{\beta}}  \label{ckkm}
\ean
for the $\sigma$-model metric of the almost complex sector. We would like to obtain this metric, as in the Calabi-Yau case, from a K\"ahler form
\ba
K \sim \log \int_X \Omega \wedge \bar{\Omega} \,.
\ea
The key equality for $G_{\alpha \bar{\beta}} \sim \partial_\alpha \partial_{\bar{\beta}}K$ to hold is the relation
\ban
\frac{\partial}{\partial z^\alpha} \Omega &=& \kappa_\alpha \Omega + \chi_\alpha   \label{defkappa}
\ean
with $\kappa_\alpha$ {\it constant}. By definition of $\chi_\alpha$,  $\tilde{\Omega}_\alpha=\frac{\partial}{\partial z^\alpha} \Omega - \chi_\alpha$ is a (3,0) form. The quotient $\kappa_\alpha = \frac{\tilde{\Omega}_\alpha}{\Omega}$ is hence well-defined. For a Calabi-Yau, $\kappa_\alpha$ must be a holomorphic function by the holomorphicity of $\Omega$ and the coordinate independence of the parameters $z^\alpha$. As a holomorphic function on a compact manifold, it must be a constant. In our more general setup, we derive this requirement from the condition that the matrices $A$, $B$, $C$ in (\ref{hodge}) be constant. In the next subsection, we will derive expressions for these constants in (\ref{starof3fcoeffs}) that depend on $\kappa_a$, and conclude that $d \kappa_a \neq 0$ is not compatible with $dA = dB = dC =0$.

Note that up to this point, the expansion of the $SU(3)$ invariant form $\Omega$ in the set $\{\alpha_A, \beta^A \}$, $\Omega = X^A \alpha_A - G_A \beta^A$, has not entered. We will need it to introduce a parametrization of $SL(3,\IC)$ structures, and argue for the metric $G_{\alpha \bar{\beta}}$ being special K\"ahler. As a first step, we want to demonstrate that the $G_A$ can be expressed as a function of the $X^A$. To this end, consider
\ba
0 &=& \int \Omega \wedge \partial_A \Omega \\
&=& \int \Omega \wedge ( \alpha_A + X^B \partial_A \alpha_B - \partial_A G_B \beta^B - G_B \partial_A \beta^B) \\
&=& G_A - X^B \partial_A G_B + X^B X^C \int \alpha_B \wedge \partial_A \alpha_C + G_B G_C \int \beta^B \wedge \partial_A \beta^C \,.
\ea
In the Calabi-Yau case, the two integrals in the final line vanish because the derivatives $\partial_A \alpha_C$, $\partial_A \beta^C$ are exact (varying the complex structure does not change the cohomology classes $[\alpha_A (X)], [\beta_A(X)]$). In the current setup, we impose the vanishing of these integrals as condition $*$\ref{h3f} on the expansion forms. The system of partial differential equations for determining $G_A$ in terms of the $X^A$, with this condition, is linear,
\ban
G_A &=& X^B \partial_A G_B \,.   \label{dhomd1}
\ean
Introducing the function $G = \frac{1}{2} G_A X^A$, such that
\ba
\partial_A G &=& G_A \,,
\ea
we see that (\ref{dhomd1}) can be rewritten as
\ba
G_A &=& X^B \partial_B G_A \,.   
\ea
The content of (\ref{dhomd1}) is hence that $G_A$ are homogenous functions of degree 1. As we have seen, they can be obtained as partial derivatives of the homogenous function of degree 2 given by $G$ as defined above.

Further, note that the RHS of (\ref{deltap}) is invariant under rescaling of $\Omega$. We can use this invariance to eliminate one of the variables, e.g. by setting $X^0=1$ away from the $X^0=0$ locus (a variation $\delta X^0=\delta$ of $\Omega$ is then implemented by the variation $\delta X^A = -\delta $, $\forall A \neq 0$). We can now introduce variables $z^\alpha$ parametrizing the variation of $\Omega$ explicitly via $z^\alpha = X^\alpha$ for $\alpha\neq 0$. As mentioned above, for these variables to parametrize variations of the $SL(3,\IC)$ structure, $\chi_\alpha$ introduced in (\ref{chi}) above must be non-zero. For a Calabi-Yau, this follows because the $\frac{1}{2}b^3 =b^{2,1}+1$ forms $\partial_A \Omega$ are linearly independent, hence span $H^{3,0} \oplus H^{2,1}$. For our more general case, we have imposed this as condition \ref{vcs}. In fact, condition \ref{vcs} is slightly stronger, and the need for it will arise in the next section.

Given this, the metric $G_{\alpha \bar{\beta}}$ in fact proves to be special K\"ahler, as in the Calabi-Yau case, with prepotential the function $G$ introduced above.

\subsubsection{The RR scalars}
In the reduction of the RR-sector, we must evaluate integrals of the form
\ba 
\int \alpha_A \wedge * \alpha_B  \,, \int \alpha_A \wedge *\beta^B\,, \int \beta^A \wedge * \beta^B\,.
\ea
This is where the coefficients $A_A^B$, $B_{AB}$, $C^{AB}$ introduced in (\ref{hodge}) come into play. Following \cite{Suzuki}, we can derive expressions for these coefficients by using the two relations
\ba
*\Omega  = -i \bar{\Omega} \,, \hspace{1cm} *\chi_\alpha = i \bar{\chi}_\alpha
\ea
(we are using conventions in which the scalar product $(\phi, \psi) = \int \phi \wedge * \psi$ is sesquilinear).\footnote{These are the conventions used e.g. in \cite{GH}. They are different from those appearing in discussions of $G$-structures, where typically one introduces a linear, rather than a conjugate linear, Hodge star operator. Under the conventions used here, no representation of $SU(3)$ is (anti) self-dual.} This first relation holds since $\Omega$ is a (3,0) form, and the second since $\chi_\alpha$ is of type (2,1) and primitive. These relations of course hold pointwise and do not require integrability of the almost complex structure. To determine the coefficients, it is convenient to undo the gauge choice $X^0=1$ and introduce the forms
\ba
\tilde{\phi}_A &=& \frac{\partial}{\partial X^A} \Omega \\
&=& \alpha_A - \partial_A G_B \beta^B + X^B \partial_A \alpha_B - G_B \partial_A \beta^B  \,.
\ea
For $A\neq0$, $\tilde{\phi}_A - \kappa_A \Omega$ is of type (2,1) with the coefficients $\kappa_A$ introduced in (\ref{defkappa}), and we define $\kappa_0$ to extend this property to all indices $A$. In the Calabi-Yau case, this (2,1) form is a sum of harmonic forms ($\alpha_A$ and $\beta^A$), and exact forms ($\partial_A \alpha_B$ and $\partial_A \beta^B$). By the commutation of the projector $\Pi^{p,q}$ on forms of definite bidegree $(p,q)$ and the projector on harmonic forms ${\cal H}$, we can drop the exact terms, obtaining $\phi_A - \kappa_A \Omega$ with 
\ba
\phi_A = \alpha_A - \partial_A G_B \,\beta^B  \,,
\ea
while maintaining the bidegree of the form. This proves crucial in deriving the precise form of the matrices $A,B,C$ needed by $\cN=2$ supersymmetry. This is why, in our more general setup, we choose to require $\phi_A - \kappa_A \Omega$ being (2,1) as condition \ref{vcs} on our expansion forms. Again by condition \ref{comp}, this (2,1) form is also primitive. Given this, $\phi_A$ satisfies the property
\ban
*\phi_A =  i \bar{\phi}_A - 2i \bar{\kappa}_A \bar{\Omega} \,. \label{keytohodge}
\ean
We can plug in the expansion (\ref{hodge}) and compute the coefficients in terms of $\kappa_A$, obtaining
\ban
C^{AB} &=& -(\im G)^{-1\, AC} (\delta^B_C -  \kappa_C X^B - \bar{\kappa}_C \bar{X}^B) \,, \nonumber\\
A_A^B &=&  C^{BC} (\re G)_{CA} + i(\kappa_A X^B - \bar{\kappa}_A \bar{X}^B) \,, \label{starof3fcoeffs}\\
B_{AB} &=& A^C_B (\re G)_{CA} - \im G_{AB} -i( \kappa_A G_B - \bar{\kappa}_A \bar{G}_B ) \,. \nonumber
\ean
As promised, for these coefficients to be constant, we must require constancy of the $\kappa_A$. Under this condition,
\ba
\kappa_A &=& \frac{\int \phi_A \wedge \bar{\Omega}}{\int \Omega \wedge \bar{\Omega}}\\
 &=& \frac{\im(G_{AB}) \bar{X}^B}{X^A \im(G_{AB}) \bar{X}^B} \,.
\ea
Substituting this relation into the above expressions for the coefficients yields the conventional result \cite{Suzuki}, and the two set of scalars assemble themselves to parametrize the quaternionic hypermultiplet scalar manifold.

\section{Expanding in eigenforms of the Laplacian} \label{s:eeL}
Our approach to this point has been to impose those conditions on our expansion forms which seem to be required for the reduction of type IIA to yield $\cN=2$ gauged supergravity -- again, we use the non-commital `seem to be required', as our approach, as we have emphasized throughout, mimics the Calabi-Yau case closely; what lies in wake when we dare to distance ourselves further from this safe haven remains to be explored. A more ambitious program would have been to justify the forms to expand in {\it ab initio}. Though concrete proposals in this direction are lacking, one natural thought is that massive eigenforms of the Laplacian should play a role in the expansion \cite{GLMW, CKT}. In the following subsection, we study the relation between our ansatz in section \ref{s:theforms} and an expansion in eigenforms of the Laplacian. In the subsequent subsection, we study how far a na\"ive approach to constructing a system satisfying the conditions of section \ref{s:theforms} based on such eigenforms takes us.

\subsection{Our conditions and the Laplacian}
On a compact manifold, the Laplacian on forms has properties close to those of a self-adjoint operator on a finite dimensional vector space. In particular (see e.g. \cite{chavel}, theorem B2), 
\vspace{0.5cm}
\begin{theorem}
The completion $L^2 A^p(M)$ of $A^p(M)$ with respect to the $L^2$ norm has an orthonormal basis $\phi_{1,p}, \phi_{2,p}, \ldots$ consisting of eigenforms of $\triangle_p$. One can order the eigenforms so that the corresponding eigenvalues $\lambda_{k,p}$ satisfy
\ba
0 \le \lambda_{1,p} \le \lambda_{2,p} \le \ldots \rightarrow \infty \,,
\ea
in particular, the multiplicities are finite.
\end{theorem}
\vspace{0.5cm}

The conditions we impose in section \ref{s:theforms} imply that our system of 2, 3, and 4-forms is closed under $d$, $d^\dagger$, and $*$. Together with the above theorem, this implies that our considerations take place within a finite number of eigenspaces of $\triangle_2$, $\triangle_3$, and $\triangle_4$. We hence have a finite basis available within which to expand our forms.

\subsection{A first attempt at constructing a set of expansion forms}
With the observation of the previous subsection, one can imagine setting out to construct a set of forms with the properties listed in section \ref{s:theforms}. We will proceed na\"{\i}vely in this subsection and obtain a set of forms that satisfy the conditions \ref{e2f} through \ref{p3f} and \ref{d23f}. One could imagine imposing condition \ref{comp} (compatibility), but condition \ref{vcs} is explicitly violated, and the reduction hence fails to yield gauged $\cN=2$ supergravity. This subsection is intended both to clarify some of the considerations in the previous sections in a more concrete setting, and to demonstrate the necessity of condition \ref{vcs}, which has not appeared in the literature previously.

We begin with a set of linearly independent 2-forms that are massive eigenforms of the Laplacian (rather than linear combinations of such) and coclosed (\cite{CKT} considers the following setup up to the proper normalization),
\ba
\triangle_2 \omega_i = m_i^2 \omega_i   \,\,,\,\,\,\,  d^\dagger \omega_i =0 \,.
\ea
With regard to the natural scalar product $(\phi,\chi) = \int \phi \wedge * \chi$, forms from different eigenspaces are orthogonal. On degenerate eigenspaces, an orthogonal basis can be introduced. Hence assume that the 2-forms $\omega_i$ form an orthogonal set. This restricts us to the metric $G_{i \bar{j}} \sim \delta_{i \bar{j}}$. We choose the normalization
\ba
|| \omega_i || &=& \frac{1}{m_i}  \,.
\ea
We introduce 4-forms according to our definition in section 2. With the normalization chosen,
\ba
\tomega^i &=&\frac{*\omega^i}{||\omega_i||^2} \\
&=& m_i^2 * \omega^i  \,.
\ea
We define a set of 3-forms via
\ban
d\omega_i &=& \alpha_i  \,\,\,,\,\,\,\,\, \beta_i = *\alpha_i.     \label{min ans}
\ean
We refer to a system of 2-, 3-, and 4-forms that satisfy the above relations as a minimal system (minimality referring to the choice of matrices $e, m, A, B, C$ relating the various forms and their Hodge duals).

Note that the 3-forms have eigenvalue $m_i^2$ with regard to $\triangle_3$.
Trivially, $\int \alpha_i \wedge \alpha_j = \int \beta_i \wedge \beta_j = 0$, and due to our choice of normalization of the 2-forms, 
\ba
\int \alpha_i \wedge \beta_j &=& \int d\omega_i \wedge *d\omega_j  \\
&=& \int \triangle_2 \omega_i \wedge * \omega_j \\
&=& \delta_{ij}  \,.
\ea
Finally, our choice of normalization of the 2-forms also guarantees the integrality of the differential of the 3-forms expanded in our set of 4-forms,
\ba
d \beta_i &=& d * \alpha_i \\
&=& d * d \omega_i \\
&=&  m_i^2 * \omega_i \\
&=&  \tomega^i \,.
\ea
This na\"{\i}ve construction hence meets the requirements \ref{e2f} through \ref{p3f} and \ref{d23f} of section \ref{s:theforms}. Condition \ref{vcs}, however, is violated. As we will now argue, this is because fixing the $*$ of the 3-forms is the moral analogue of compactifying on a Calabi-Yau with rigid complex structure. To see this, consider again
\ba
*\Omega  = -i \bar{\Omega} \,.
\ea
In section \ref{s:rms}, we use this condition to determine the matrices $A,B,C$. Since these matrices are fixed in the minimal setup of this section, this condition instead allows us to solve for $G_i$, and yields
\ba
G_i&=& iX^i  \,,
\ea
such that $\Omega = X^i (\alpha_i - i \!*\!\!\alpha_i)$. The variation of $\Omega$ with regard to $X^i$ now clearly does not contain a $(2,1)$ piece, hence does not correspond to a variation of $SL(3,\IC)$ structure. Without this condition, the reduction fails to assemble itself into a quaternionic sector.

\subsection{The scope of minimality}
We witnessed in the previous subsection that the minimal system fails to satisfy the complete set of constraints required to yield the desired reduction. In the form (\ref{min ans}), the minimal system is easy to identify. However, all ans\"atze related to (\ref{min ans}) via a symplectomorphism are equivalent and will equally fail. In view of this, we consider in this final subsection what conditions the matrices $m,e, A,B,C$ must satisfy for our system of forms to not be equivalent to the minimal system.

To transform a given system with 
\ban
d\omega_i &=& m_{ij} \alpha_j + e_{i j} \beta_j     \label{nonmin}
\ean
to a minimal one, we need to find a symplectomorphism 
\ba
\left(\begin{array}{c}\alpha' \\\beta'\end{array}\right) &=&
\cM \left(\begin{array}{c}\alpha \\\beta\end{array}\right)
\ea
such that
\ban
(\cN \alpha')_i &=& m_{ij} \alpha_j + e_{i j} \beta_j \,, \nn\\
\beta'_i &=& * \alpha'_i \,,  \label{condonM}
\ean
with $\cN$ a real invertible matrix. We can then introduce a new set of two forms
\ba
\omega'_i &=& (\cN^{-1} \omega)_i \,,
\ea
thus reexpressing (\ref{nonmin}) in minimal form
\ba
d \omega_i' = \alpha'_i  \,\,\,\,,\,\,\,\,\,\, \beta'_i = *\alpha'_i   \,.
\ea
When does such an $\cM$ exist? By (\ref{condonM}), 
\ba
\cM&=& \left(\begin{array}{cc}\cN^{-1} m & \cN^{-1} e \\\cN^{-1} (mA +eC) & \cN^{-1} (mB - eA)\end{array}\right)  \,,
\ea
yielding the conditions
\ba \small
\cN^{-1} \left(\begin{array}{cc}  me^T -em^T  &  m B m^T - e C e^T - m A e^T - e A m^T   \\  -\left[ m B m^T - e C e^T - m A e^T - e A m^T \right]  &  (mA + eC)(mB-eA)^T  - transp.   \end{array}\right)(\cN^{-1})^{T}\\
=\left(\begin{array}{cc}0 & 1 \\-1 & 0\end{array}\right) \,,
\ea
where we have written $\cN$ for $\cN \otimes \tiny\left(\begin{array}{cc}1 & 0 \\0 & 1\end{array}\right)$ for notational simplicity.
The first condition (position (1,1) in the matrix) is just (\ref{nilpcond}), required by $d^2=0$. The condition
\ban
-m B m^T + e C e^T + m A e^T + e A m^T  = \cN \cN^T  \label{c12}
\ean
can be used to determine $\cN$. A solution exists, since the matrix on the LHS is symmetric, but for $\cN$ to be real, the eigenvalues of this matrix must all be positive. Finally, we obtain the condition that the matrix
\ban
(mB-eA) (mA + eC)^T    \label{c22}
\ean
be symmetric.

To recapitulate, if the matrices $e,m,A,B,C$ are such that these two conditions are satisfied, our system of expansion forms (\ref{nonmin}) is equivalent to a minimal system and hence not suitable as a starting point for the reduction. Note finally that particularly in this final section, we have been treating the matrices $A,B,C$ as an input. Hopefully, a deeper understanding of the type of $SU(3)$ reduction discussed in this paper will have an intrinsic definition of the $X^A$ and $G_A$ of equation (\ref{expinvf}) as a starting point, and these matrices will then follow from (\ref{starof3fcoeffs}), as in the Calabi-Yau case.

\section*{Acknowledgements}
We would like to thank A.~Tomasiello for initial collaboration on this project and many useful discussions. We gratefully acknowledge useful conversations with Jan de Boer, Sheer El-Showk, Mariana Gra\~na, Shamit Kachru, Kostas Skenderis, Marika Taylor, and Daniel Waldram. 
AK would like to thank the Aspen Institute for Physics, where part of this work was performed. RM would like to thank the Institute
for Mathematical Sciences, Imperial College for hospitality.
The work of AK was supported by Stichting FOM. RM is supported in part by an RTN contract MRTN-CT-2004-005104  and by ANR
grant BLAN06-3-137168.

\appendix
\section{A variation on conditions 3 and 4}
Note that in the reduction, the $*$ of the 3-forms $\alpha_A, \beta^A$ always appears in integrals of the form
\ba 
\int \alpha_A \wedge * \alpha_B  \,, \int \alpha_A \wedge *\beta^B\,, \int \beta^A \wedge * \beta^B \,.
\ea
Hence, allowing additional terms in the expansion of $*\alpha_A, *\beta^A$ that vanish upon integration does not alter the reduction. Given our condition $*$\ref{h3f}, we can hence live with
\ban
* \alpha_A &=& A_A^B \alpha_B + B_{AB} \beta^B + D_{1A}^{BC} \partial_B \alpha_C  + D_{2A}^{BC} \partial_B \beta_C \,,  \nn \\
*\beta^A &=& C^{AB} \alpha_B - A^A_B \beta^B +D_{3A}^{BC} \partial_B \alpha_C  + D_{4A}^{BC} \partial_B \beta_C\,,
\ean
rather than (\ref{hodge}) in condition \ref{p3f}.
If we further demand $d\kappa_A = 0$, rather than derive this condition from the constancy of the matrices $A,B,C$ as in the text,
we again obtain the expressions (\ref{starof3fcoeffs}) for the matrices $A,B,C$, now by integrating the relation 
\ba
*\tilde{\phi}_A =  i \bar{\tilde{\phi}}_A - 2i \bar{\kappa}_A \bar{\Omega} \,,
\ea
i.e. (\ref{keytohodge}) with $\phi_A$ replaced by $\tilde{\phi}_A$,
against $\alpha_B$ and $\beta^B$. We then no longer need to introduce $\phi_A$ as in (\ref{bidegreeconstr}) of condition \ref{vcs}, and can instead demand that $\tilde{\phi}$ directly be of type (2,1). The price we pay, aside from having to impose $d\kappa_A =0$ by hand, is that our system of forms is no longer closed under $*$.

\end{document}